# Growth of textured quasicrystalline phase in Ti-Ni-Zr films prepared by pulsed laser deposition

**V. Brien[1], A. Dauscher[2], P. Weisbecker[1], F. Machizaud[1]**

1 Laboratoire de Science et Génie des Matériaux et de Métallurgie, UMR 7584, CNRS-INPL-UHP, Parc de Saurupt, ENSMN, 54042 NANCY Cedex, FRANCE

2 Laboratoire de Physique des Matériaux, UMR 7556, CNRS-INPL-UHP, Parc de Saurupt, ENSMN, 54042 NANCY Cedex, FRANCE

For correspondence: V. Brien, E-mail: brien@mines.u-nancy.fr, Fax: 00.33.3.83.57.63.00

"This paper is dedicated to Dr André Simon on the occasion of his 60[th] birthday"

## ABSTRACT

**The preparation in thin film form of the known icosahedral phase in Ti-Ni-Zr bulk alloys has been investigated as a function of substrate temperature. Films were deposited by Pulsed Laser Deposition on sapphire substrates at temperatures ranging from room temperature to 350°C. Morphological and structural modifications have been followed by grazing incidence and $\theta$-$2\theta$ X-ray diffraction, transmission electron diffraction and imaging. Chemical composition has been analysed by Electron Probe Micro-Analysis. The in-depth variation of composition has been studied by Secondary Neutral Mass Spectroscopy. We show that Pulsed Laser Deposition at 275°C makes the formation of a 1 μm thick film of Ti-Ni-Zr quasicrystalline textured nanocrystallites possible.**

## INTRODUCTION

Quasicrystals are known to exhibit a certain range of properties as new as varied considering they are made of metallic atoms. Some icosahedral quasicrystalline Ti-based alloys or related phases have been shown by Kelton et al. and Nicula et al. [1,2] to be able to reversibly store large amounts of hydrogen. As a matter of fact, the large amount of tetrahedral sites in the structure combined with the natural affinity of Ti and Zr atoms to hydrogen makes the Ti-Ni-Zr quasicrystalline phase an excellent candidate for constitutive materials in batteries. The phase can actually absorb up to two hydrogen atoms for each metallic atom [3]. This performance makes this new phase a tangible alternative as a hydrogen storage material. In order to be able to use these material in this goal the feasability of processing films must be proven as the optimization of the storage in such solid batteries needs the maximisation of the surface/volume ratio of the material [4]. Pulsed Laser Deposition (PLD) was chosen as the preparation method because it has shown to be a powerful method to prepare thin films of virtually any material, from pure elements to multi-component compounds [5]. Stoichiometry of the target is generally reproduced in the deposited film. In contrast to conventional deposition techniques, PLD provides rather unusual growth conditions: very high supersaturation, bombardment by energetic particles and extremely high quenching rates of the condensing atoms. Consequently the built-up of the film takes place far from equilibrium conditions, leading to the formation of metastable film structures. Significant differences in structure and microstructure of simple metallic alloys films prepared by PLD have been found as compared with films prepared by more conventionnal techniques [6,7]. Actually, very few papers deal with the preparation of quasicrystals by laser ablation. Indeed, preparation of quasicrystalline films has most commonly been tackled using other techniques. For instance, post-annealing of sequential electron gun deposition of constitutive elements of the alloy was used to prepare Al-Co films [8], Al-Cu-Fe films [9-11] or Al-Cu-Mg films [12]. Physical vapor deposition was also used [13] to make Al-Cu-Fe films. By laser ablation, Ichikawa et al. [14] are the only one to get a pure quasicrystalline film in the Al-Pd-Mn system, while Teghil et al. [15] obtained a mixture of the icosahedral $\varphi$ phase and the crystalline $\beta$ phase of the Al-Cu-Fe system. In this paper, we report for the first time on the preparation of quasicrystalline Ti-Ni-Zr thin film by PLD. Attention was focused on the optimisation of the substrate deposition temperature. Particular care was taken to characterize the chemical composition, and the microstructure changes.

## 1. EXPERIMENTAL PROCEDURE

### 1.1 Synthesis

A pulsed Nd:YAG (Qantel, YG 571C) laser was used for the ablation process, working at the fundamental wavelength (1064 nm), a repetition rate of 10 Hz and pulse durations of 10 ns. The laser beam impinged at 30° with regard to the plane of the target that was placed in a vertical position. It was located at a distance of 45 mm from the substrate kept at the temperature $T_s$ by simple Joule effect of a small resistance located behind (figure 1). The energy of the outcoming beam was monitored by an energymeter, allowing the estimation of the laser fluence used: 72 J/cm$^2$.

Experiments were conducted in a high vacuum chamber under an initial base pressure of $10^{-7}$-$10^{-8}$ mbar stabilizing at a value of one decade order less after few minutes of ablation.

The targets used for the pulsed laser deposition were obtained from cutting Ti-Ni-Zr ingots further polished with SiC 1200 paper grid. The ingots were prepared by RF melting high purity metals under helium atmosphere in an induction furnace. Oxidation of Ti and Zr were prevented as far as it was possible by preparing them in an argon glove box. The targets presented as disks of 18 mm in diameter and 4 mm in thickness. The bulk materials were characterized by electron probe micro-analysis (EPMA) on a CAMECA SX 50 and Scanning Electron Microscopy (SEM) with a PHILIPS XL30G in order to check the composition of the ingots and more particularly that the size of the different phases present in the bulk target were far smaller than the used laser spot size. The same target was used for the preparation of several consecutive films, provided a standard polishing was performed prior to any film deposition. This polishing ensured among others the elimination of surface roughness of the target that usually produces non wanted droplets on the film. Following the same goal, combination of a non-linear motion of the laser beam through a computerized control of step by step motors [16] and a rotation of the target (11 r/min) was used to renew the irradiated area and to avoid crater formation.

Sapphire substrates oriented along their [0001] direction used as received were chosen as a substrate material due to their chemical inertness and role as diffusion barrier. The obtained thick films were all strongly adherent to their substrate and mirror like. Typical deposition rates obtained were 1-8 Å/min.

The films were not further annealed as destructive oxidation systematically occurred during the process.

### 1.2 Characterization

The global composition of the films was characterized by EPMA when thick enough (no substrate interference). Chemical depth profiles were obtained by Secondary Neutral Mass Spectroscopy (SNMS). All the chemical compositions are given within the experimental error of 1 atomic %. The thicknesses were estimated by performing profilometry on the SNMS craters. Grazing incidence X-Ray Diffraction (GXRD) using a Co$K_{\alpha1}$ radiation and a curved detector (INEL CPS120) was performed at 0.5° incidence for the identification of the phases. GXRD is an excellent technique to inform about the nature and the structure of the phases formed in thin films. It is however extremely crucial to specify the geometrical configurations of the atomic planes that diffract under such incidences. Indeed, as the curved detector collects X-rays spanning 120° $2\theta$ range and the incidence angle is fixed to 0.5°, the atomic planes susceptible to be detected when in Bragg conditions are the ones, on one hand whose zone axis is perpendicular to the incident plane and parallel to the goniometer axis and on the other hand that make an angle of 0 to 60° with the surface. Such a diffraction geometry implies that a peak measured on the pattern at $2\theta_0$ indicates the presence in the film of grains such orientated that the related family of planes are inclined at an angle of $\theta_0$ - 0.5 ° towards the surface. Carbon coated copper electron microscopy grids were systematically installed just next to the substrate on the substrate holder inside the deposition chamber. Transmission Electron Microscopy (TEM) observations were carried out on the material deposited on these grids or by simple levy on the film with a cutter for the thick 1 μm film. A PHILIPS CM200 with accelerating voltage of 200kV has been used. This thickest film was also characterized by standard Bragg-Brentano classic $\theta$-$2\theta$ diffraction using the Co$K_{\alpha1}$ radiation and by 4 circles diffraction using the Co$K_\alpha$ radiation in order to describe the observed texture. To do so, four pole figures were recorded. One recalls a pole figure of type [hkl] represents the angular distribution of intensities of the reflections of this [hkl] Bragg family on a stereographic projection. In concrete terms, the sample was rotated through the angle $\varphi$ about its normal direction and through the angle $\chi$ about the intersection of the reflecting plane and the sample plane. The origin of the angle $\chi$ is parallel to the normal direction of the film. The data were corrected and analysed using software Texeval™ from Bruker axs. Iso-intensities lines have been chosen such that the most intense zones appear on every pole figure. Error on measurement of $\chi$ ($\chi$ being then the radial angle on the pole figures) was checked on a monocrystalline Si sample and found to be around 1°.

## 2. RESULTS AND DISCUSSION
### 2.1 Chemical congruence

The study of the chemical transfer between the targets used and the films obtained revealed the ternary Ti-Ni-Zr system does unfortunately not transfer the chemistry as faithfully as expected. Two points have been made clear.

To get the $Ti_{41.5}Ni_{17}Zr_{41.5}$ desired composition which was identified by Electron Dispersion Spectroscopy by Kelton et al. in [17] to be the icosahedral phase nominal composition, a target of $Ti_{45}Ni_{17}Zr_{38}$ composition was found to give the best results at room temperature. Indeed, a 1 μm thick film of Ti-Ni-Zr was obtained keeping the substrate temperature at 25°C. Its atomic composition was measured by EPMA and found to be: $Ti_{42}Ni_{18}Zr_{40}$ and fairly homogeneous on the whole surface deposited.

Since the measured composition by EPMA represents the integration of around 1μm$^3$ volume, depth chemical profiling has been investigated by SNMS in the same 1 μm film and is shown in figure 2. From this profile it is possible to follow the chemical transfer chronology. Close to the interface with the substrate, the stoichiometry transfer from target to substrate is far from being achieved: larger amounts of Zr to the detriment of those of Ni and Ti are observed. The wanted stoichiometry $Ti_{41.5}Ni_{17}Zr_{41.5}$ (different from the target) can only be observed in the 0.8-0.5 μm range of figure 2, after a transition regime that lasts roughly a duration of 30 min. This duration is much larger than what is necessary to erode the whole target surface at one time, meaning that the Zr enrichment in the film is certainly not only linked to Zr target surface segregation but also to preferential erosion. The corresponding depletion in Ti may rise from re-sputtering from the film surface. So, the wanted stoichiometry is only reached after having deposited around 0.2 µm and presents the most interesting transfer regime during 0.3 µm (0.8 to 0.5 µm in figure 2). Provided a pre-ablation of 30 min is assured prior to any film growth, the first 0.3 µm of deposited material will have the right composition within 1at.% and the following layers within 2at.%, presenting such a slight chemical gradient as figure 2 shows it (range (0-0.5µm).

## 2.2 Dependance of the nature and structure of the films versus the substrate temperature

In order to be able to practise electron microscopy the films deposited have been designed to be around 50 nm thick. The structures of the films were all complementarily analysed by GXRD and TEM.

### 2.2.1 X-ray diffraction results

The X-ray diffraction pattern of the film deposited at room temperature presented in figure 3 shows a broad maximum which characterizes an amorphous structure. To free from this non wanted amorphous phase, more energy was brought to the substrate seeking for a crystallisation of the deposit. A series of film depositions has been prepared heating the sapphire substrate during the laser ablation process. Chosen temperatures were 160°C, 240°C, 295°C and 350°C, and results of GXRD are respectively shown in figure 4. When identified, the indexation of the peaks has been placed on the figures. The system of indexation by Cahn et al. [18] was used for the quasicrystalline phase. The reader will find in the appendix a correspondence table for the most intense peaks with the Bancel et al. [19] system of indexation. Details on the approximant W phase found to appear in lower temperatures can be found in [20].

Figure 4 presents the GXRD pattern of the film prepared at 295°C. It exhibits reflections that could be identified as the three main peaks of the Ti-Ni-Zr icosahedral structure: the (2/3 0/0 1/2), the (2/4 0/0 0/0) and the (4/6 0/0 0/0) reflections. Their corresponding experimental $Q_{//}$ are listed in table 1. One recalls $Q_{//} = (2\pi/a_{6D})((N+M\tau)/(2(2+\tau)))^{1/2} = (2\pi/a_{6D}) f(N,M)$ with the gold number $\tau = (1+\sqrt{5})/2$ and $Q_{//} = 4\pi\sin\theta/\lambda$, setting $f(N,M) = ((N+M\tau)/(2(2+\tau)))^{1/2}$ leading to:

$$a_{6D} = (2\pi/Q_{//}) \; f(N,M) \qquad (1)$$

The six dimensional parameter value was found to be $a_{6D}$ = 0.741 ± 0.001 nm (quasilattice $a_q$ = 0.52 ± 0.001nm). The increase of the lattice parameter of 0.6 % compared to the values obtained by high resolution X-ray diffraction (synchrotron) found in literature ([2], $a_{6D}$ = 0.7368 ± 4.10$^{-4}$ nm, $a_q$ = 0.5210 ± 4.10$^{-4}$ nm) can probably be explained by oxygen absorption of the lattice (SNMS dosage of oxygen actually gives 1.5 at.%) or by a phenomenon of lattice expansion usually observed for intermetallic compound films deposited by the PLD technique because of very intense condensation speeds [21]. The comparison of the relative intensities of the icosahedral peaks and the ones usually obtained in literature [2, 3, 22-26] for the bulk phase informs that the icosahedral grains are not 100% randomly distributed. Relative intensities of all the peaks show the icosahedral phase is the main phase present in that film. The shoulder of the 42° peak and the 51° peak witness the film also contains a non wanted second phase which becomes predominant at higher temperature (350°C).

At lower substrate temperature, i.e. 240°C, deconvolution was necessary to understand the signal obtained by GXRD around 2$\theta$ = 42° (figure 4). As figure 5 shows it the signal could be decomposed in 3 peaks. The details of all the peaks of this pattern namely position, full width at half maximum (FWHM), indexation, deduced parameter and size of the diffracting particles calculated thanks to Scherrer law are given in table 2. The most intense peak of the pattern does not correspond to any phase of the ICDD database (Powder Diffraction file$^{TM}$ Release 2000) or to any peak of a phase usually met in works dealing with the preparation of Ti-Ni-Zr icosahedral phase. The only possible match is the (2/3 0/0 1/2) reflection of the icosahedral structure. Six dimensional parameter then obtained is $a_{6D} = 0.745 \pm 0.002$ nm ($a_q = 0.527 \pm 0.002$ nm). Missing of the other peaks of the icosahedral structure linked to this one suggests the grains of this domain (labelled 1) are strongly textured (to be dealt further). The average size of the crystallites of this domain is 16 nm. The analysis of the other peaks evidences that the film contains also a second domain (labelled 2) made of smaller imperfect quasi-crystalline crystallites (see table 1). Their average size is around 4 nm.

At 160°C, the film seems to be formed by a mixture of the icosahedral phase and of another phase producing the spreading of the large complex peak centred around $2\theta = 42°$. This enlargement could be due to residual presence of the amorphous phase met at room temperature or to the presence of another phase. One probable phase is the known 1/1 approximant: the *W* phase [20]. Indeed, the (530), (532) and (006) peaks of the *W* phase placed on the pattern correspond rather well to the presence of the different shoulders on the peak. It is however rather difficult to conclude. Indeed, the *W* phase usually appears at higher temperature than the icosahedral one during annealing treatments of bulk material [27].

The patterns attesting the presence of the icosahedral phase display different sets of peaks of the structure versus the temperature substrate. The grains are therefore differently orientated against the substrate temperature and suggest that different textures of this phase might be obtainable by changing the temperature.

**2.2.2 TEM results**

TEM diffraction and TEM imaging in dark field were also performed in order to assess the size and geometry of the grains. Figure 6 shows a pattern obtained for the sample prepared at 240°C and figure 7 for the one deposited at 295°C. Whatever the temperature of the sample, the diffraction patterns exhibit well defined rings witnessing all the films are made of small grains, and present strong radial but symmetric heterogeneities so that the rings intensities contrasts make symmetric arcs. Such arcs are typical of the presence of textures and are present in all diffraction patterns. In the same way, rings of the icosahedral structure are systematically present, but the distribution of intensity on the rings is different from one pattern to another and depends on the temperature $T_s$.

Since X-ray diffraction results of the films prepared at 240°C and 295°C show these temperatures are the most appropriate to maximize the volume of the icosahedral phase, only TEM data from samples obtained at these temperatures are presented (figure 6a and b). Again, Cahn et al. [18] indexation was used and placed on the figures.

Table 1 lists the values of $Q_{//}$ obtained from the measurements of the diameter of the rings on either pattern (240°C or 295°C) since they exhibit the same inter-reticular distances. Parameter found was: $a_{6D} = 0.72 \pm 0.04$ nm ($a_q = 0.52 \pm 0.04$ nm). The difference of 3% with what is obtained in GXRD (cf. 2.2.1) for the films is of the same order of the 5% error coming from the TEM technique itself despite calibration.

On top of the rings belonging to the icosahedral structure as figure 6a presents it, diffuse intensity can be noticed on the TEM pattern made on the 240°C sample around rings (1/4 0/1 0/0), (2/3 0/0 1/2) and (2/4 0/0 0/0) corresponding to the identified crystallites of domain 2 (see analysis of GXRD data). At 295°C, the icosahedral domain 2 has disappeared.

The systematic dark field patterns performed localizing an aperture on an arc of the fifth diffused paraxial ring (2/4 0/0 0/0) on the different samples allows an estimation of the sizes of the crystallites. They are listed in table 3 versus the substrate temperature $T_s$, smaller and bigger values are indicated. Sizes distribution seems to be simple and the average size of the particles is apparently in the middle of the ranges. Size homogeneity was, however, checked by systematic measurements on several images. The sizes can reach up to 10 or even 20 nm for $T_s = 295°C$.

**2.3 Growth of a thick icosahedral textured Ti-Ni-Zr film sample**

As presented above, a substrate temperature of 240°C leads to the formation of the icosahedral phase divided up in two domains. One of the domains is made of textured crystallites, the other one is made of smaller non textured grains. On the other hand, 295°C has been shown to lead to a mixture of the icosahedral phase and a high temperature crystalline phase. Consequently, 275°C was the temperature chosen for the preparation of a monophased quasicrystalline film hoping a possible texture. To get a complete description of the expected texture of the film, its thickness was goaled around 1 μm. Two hours of ablation were necessary to get the right thickness.

The chemical composition of the film obtained by EPMA gives: $Ti_{42.9}Ni_{15.6}Zr_{41.5}$, to be compared to $Ti_{41.5}Ni_{17}Zr_{41.5}$ the composition found for the quasi-crystalline phase in bulk by other authors [17].

The Bragg-Brentano, GXRD and 4 circles diffraction patterns performed on the thick film respectively presented in figures 8a, 8b and 8c bring complementary structural information that are all consistent. The Bragg-Brentano diffraction pattern exhibits peaks that could be indexed as (2/3 0/0 1/2) and (4/6 0/0 1/4) of the icosahedral structure. The GXRD pattern of the same film exhibits other peaks of the icosahedral structure like (2/4 0/0 0/0) and (4/6 0/0 0/0) which are the most intense. The presence of these different reflections on the two patterns attests of the presence of a texture in the film. To precise this texture, 22 scans were recorded at different $\chi$ on the 4 circles diffractometer and added in order to cumulate and vizualise all the possible Bragg reflections of the film on a same pattern. The acquisition of the 22 cumulated scans were done with $\chi$ ranging from 0 to 55° assuring a $\varphi$ rotation of the sample at 600 rotations per minute (figure 8c). Such an acquisition allows to visualize all the possible Bragg diffractions of all the crystallographic planes present in the film for Bragg angles located between $2\theta = 30°$ and $2\theta = 96°$. The obtained pattern exhibits a set of reflections which can be attributed to a primitive (P-type) hypercubic lattice of an icosahedral phase (icosahedral group $m\overline{3}\overline{5}$). Indexation of all peaks has been inserted in the pattern. The lattice parameter refinement has been performed and gives $a_{6D} = 0.744 \pm 0.001$ nm ($a_q = 0.526 \pm 0.001$ nm).

The texture was specified by recording pole figures for different planes (figure 9): chosen Bragg peaks are on one hand $2\theta = 42°$ and 92.3° on the other hand 44° and 76° respectively corresponding to the main reflections of figures 8a and 8b.

According to the indexation performed above, the 42° and 92.3° peaks come from the same family of planes perpendicular to fivefold axes, indeed the 92.3° peak corresponds to a second order reflection. Figure 9a exhibits a maximum of diffraction around $\chi = 0°$ in the shape of a half-circle located at $\chi = 6.4°$. This family gives very intense maxima at $\chi = 56°$ and $\chi = 68.5°$. Measurements on the pole figure (figure 9a and 9b) of the angle $\Delta\chi$ between the most intense domains lead to the following values: $62.1 \pm 2°$ and $62.4 \pm 2°$ (68.5° - 6.4° or 56° + 6.4°). This is to be compared with 63.4° which is the angle between two fivefold axes in the icosahedral structure [28-29]. Comparison of figures 9a and 9b shows they present exactly the same aspect with an attenuation in intensity from 8a to 8b. The small half-circle is also positioned exactly at $\chi = 6.4°$ and despite their weakness traces of diffraction intensity could be recorded at the position corresponding to the maxima of the half-circle located at 56° and 62.4°. It unambiguously confirms these two peaks are related to the same family of planes. A compression of the parameter of around 1% is also noticed when calculating it from the position of the (2/3 0/0 1/2) reflection when scanning at $\chi = 6°$ and $\chi = 56°$: one finds 0.748 nm at 6° and 0.738 nm at 56°.

A similar analysis was performed on the two most intense peaks appearing on figure 8b (GXRD). Figure 9c and 9d present the pole figures of the 44° and 76° peaks. The identical distribution of intensity on the two pole figures confirms they concern parallel planes. Indeed, the indexation informs the two peaks are Bragg peaks of families od planes perpendicular to twofold axes and are so linked by a $\tau$ deflation. Similarly, $88.5 \pm 2°$ and $89.5 \pm 2°$ (24.5° + 64° or 37.5° + 52°) are the angles measured between respectively the rings labelled 1 and 4 of figure 9c, and between the rings labelled 2 and 3. These values are to be compared with 90°, the angle between twofold axes in the icosahedral structure.

On the top of all this analysis, it was possible to check that the angles between what is deduced as reflections linked to planes perpendicular to fivefold and twofold axes make really angles fully compatible with the icosahedral structure: indeed 31.5 ± 2° (56° - 24.5°) and 31.0 ± 2° (68.5° -37.5°) are measured. This is to be compared with the angles between fivefold and twofold axes: 31.7°. In the same way, it has been carefully checked the family of planes (2/4 0/2 0/0) perpendicular to three fold axes gives compatible angles (not presented here). Such a texture study shows the planes perpendicular to the fivefold axes of the icosahedral Ti-Ni-Zr phase grow nearly parallel to the surface film (inclination of 6.4°). The texture observed here could be described as a non-homogeneous fibre tilted of 6.4° with respenct to the normal of the film. Standard fibre textures are very often met in growth of thin films [30-32]. The angle of the maximum of the texture with respect to the normal of the film and probably also the non-homogeneous distribution of intensities could have been produced by the inclination of the symmetry centre of the plasma plume inside the laser chamber with regard to the normal of the substrate [33].

Under the light of this four circles diffraction investigation, it is rather trivial to understand why only reflections from planes perpendicular to fivefold axes are visible in figure 8a (Bragg-Brentano diffraction) and that only the ones perpendicular to twofold axes in figure 8b (GXRD). Indeed, geometrical configurations of GXRD and Bragg-Brentano diffraction techniques make that in Bragg-Brentano diffraction only the Bragg planes parallel to the film surface diffract towards the detector and that in GXRD only the planes making an angle with the film surface in the range 0 and 60° and equal to their Bragg angle diffract towards the curved detector. So, as figure 9 shows it: the [2/3 0/0 1/2] and the [4/6 0/0 2/4] reflections diffract near $\chi = 0°$, explaining then why they will be detected on a Bragg-Brentano pattern. In the same way, the [2/4 0/0 0/0] reflection is present on the GXRD pattern because the film contains related planes located around its Bragg angle 22°, as the ring labelled 1 and localised around 24.5° on figure 9c attests it. The [4/6 0/ 0 0/0] is also detected on GXRD pattern because the corresponding diffraction planes are found on figure 9d around $\chi = 37.5°$ (Bragg angle = 38°).

To complete this study, TEM has been performed (figure 10). Most parts of the film produce a diffraction pattern made of dotted rings. Imaging informs us that the nanocrystallites are slightly bigger than in the films prepared at other temperatures (13 nm ± 4 nm, figure 10a). Some rare larger grains have also been noticed producing a line of sharp Bragg spots, well aligned in a sequence of distances to the centre of the pattern in $\tau$ ratios (figure 10b). The thickness and configuration of those grains did unfortunately not allow to get a pattern along a zone axis of the structure. This diffraction line can be a systematic line from a fivefold electron diffraction pattern of an icosahedral structure or a systematic line from a pseudo-fivefold diffraction pattern of an approximant to the Ti-Ni-Zr quasicrystal [23].

**CONCLUSIONS**

This is the first report on the preparation of a film of the quasicrystalline Ti-Ni-Zr phase. Successful preparation was reached using the PLD technique deposition. Experimental conditions of preparation were found and are such the films do not need any further thermal treatment. The one micron synthetized film presents the $Ti_{41.5}Ni_{17}Zr_{41.5}$ icosahedral phase composition known in bulk material to within about 2 atomic %. It is made of icosahedral nanosized grains presenting a texture along a fivefold axis slightly tilted (≈ 6°) from the normal to the film. The atomic planes have also been shown to be stressed in the directions perpendicular to the growth, as usually observed in such elaborated films. On the other hand, rare but much bigger approximant or icosahedral grains than the nanosized ones have been noticed in the film. The grown films were all strongly adherent to their substrate thanks to small sized grains. Investigations performed suggest also different kind of textures might be obtained by changing the substrate temperature during deposition. The confirmation of such a promise would be extremely interesting as hydrogen storage is the main promising industrial application of the phase. One could then study whether energy storage depends on the films textures.

**Acknowledgements:**

The authors would like to thank J.-M. Dubois for having suggested this study. They also would like to thank J.-P. Haeussler and J.-B. Ledeuil for their assistance in microprobe measurements and S. Weber for SNMS experiments. We are most grateful to F. Brillancourt from LCTMR laboratory (CNRS France) for her help to prepare the ingots and to A. Percheron (LCTMR) for her constant interest in this work.

**Figure and table captions :**

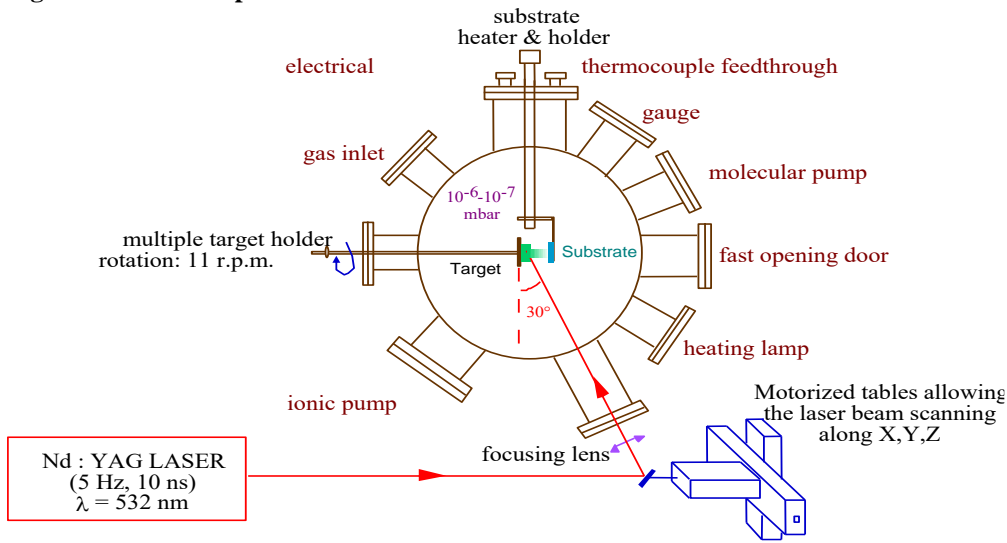

Figure 1: Schematic diagram of the pulsed laser deposition set-up

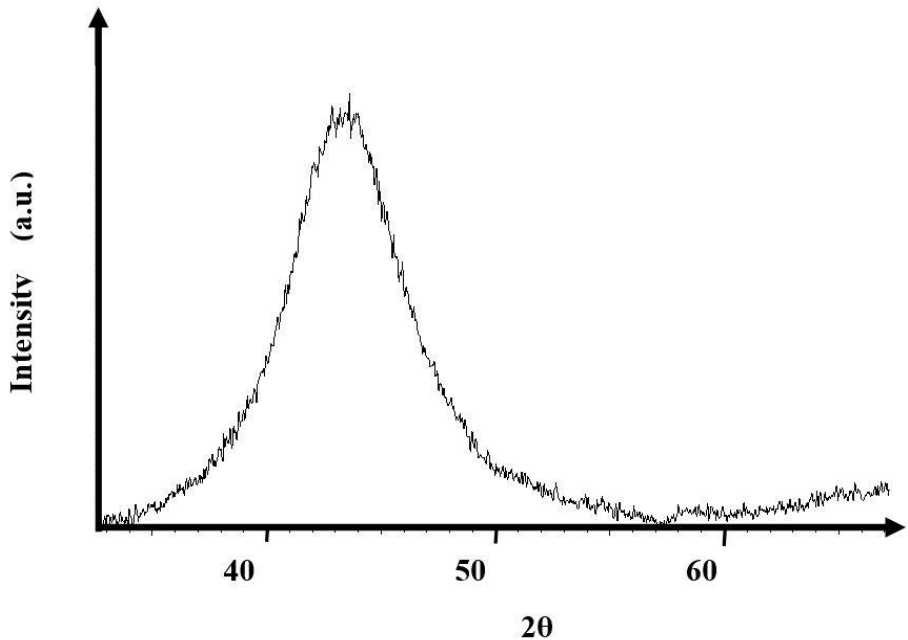

Figure 2: In-depth ternary chemical profile obtained by SNMS on a 1μm Ti-Ni-Zr film deposited at room temperature from a $Ti_{45}Ni_{17}Zr_{38}$ ingot imaging the chemical transfer chronology. Substrate is on the right, above 1μm.

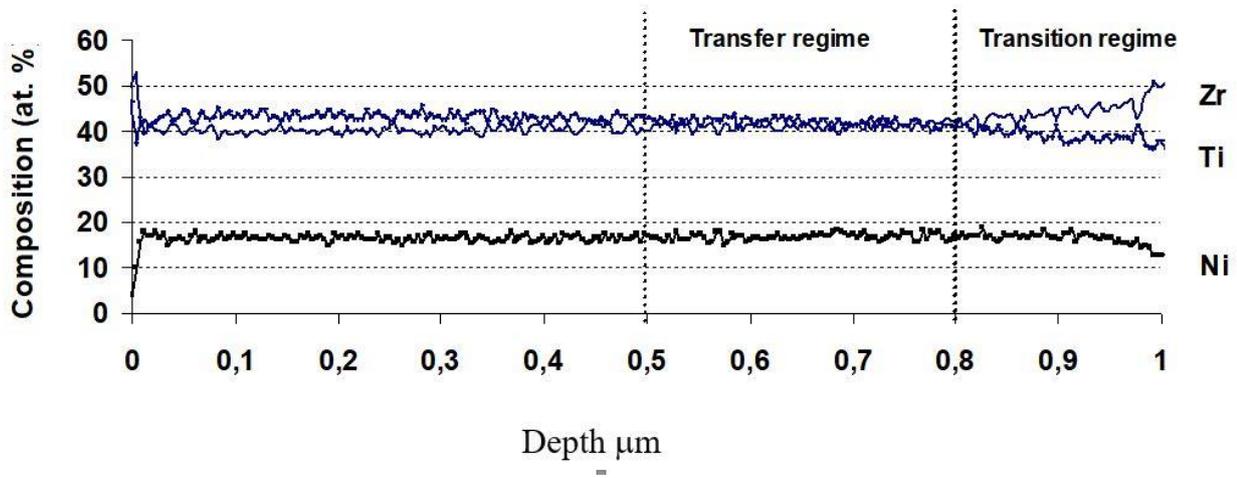

Figure 3 : Grazing incidence (0.5°) X-ray diffraction pattern of the Ti-Ni-Zr film deposited at room temperature (λ = 0.178897 nm). Background has been removed.

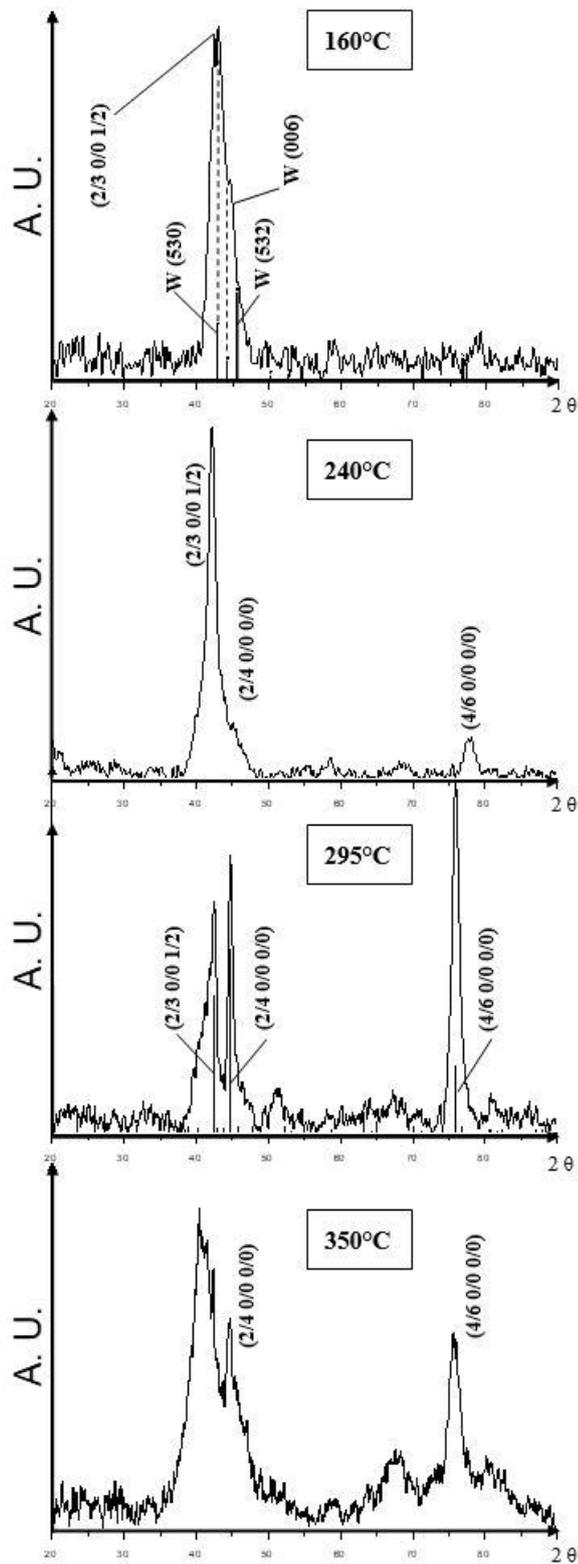

Figure 4: Set of grazing incidence X-ray diffraction patterns of the Ti-Ni-Zr films deposited at various temperatures, 160°C ≤ T ≤ 350°C. The vertical lines indicate the position of the main peaks of the quasicrystalline phase and the W phase. See deconvolution of the 2 θ = 42° peak of the 240°C X-ray pattern in figure 5 and table 1.

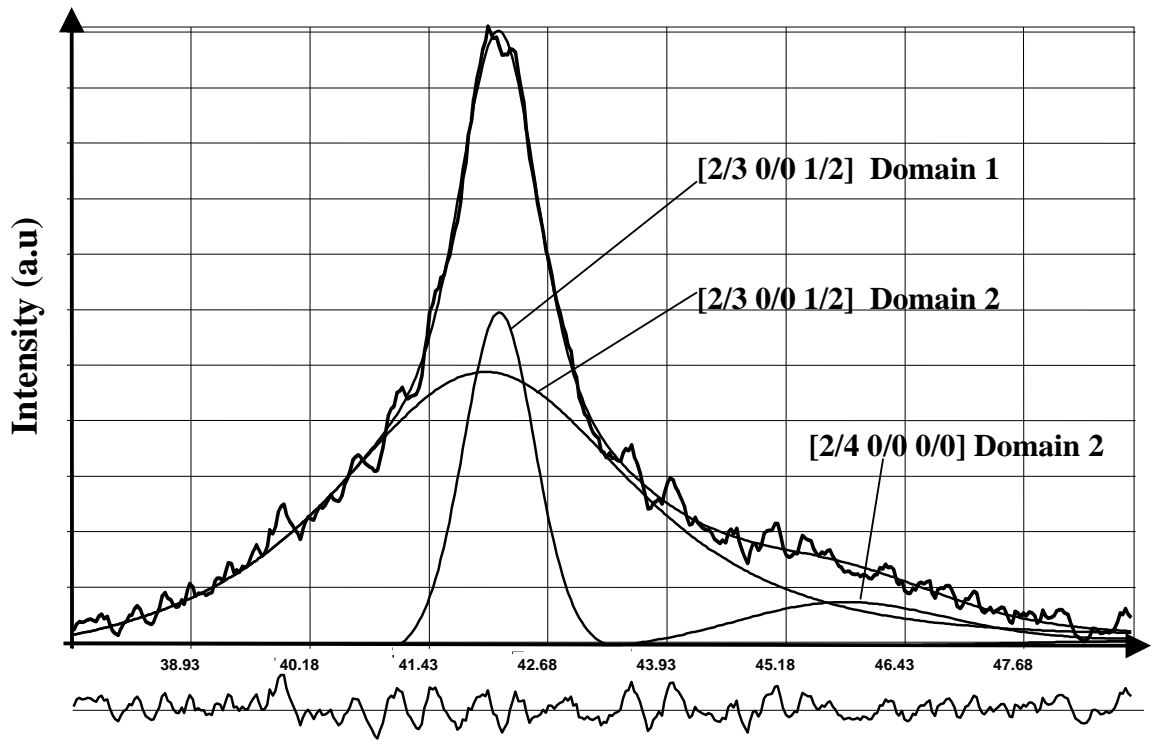

Figure 5 : Detail of deconvolution using winfit software* of the 42° peak of the GXRD pattern obtained on the film prepared at 240°C. The lower curve is the difference between the experimental data and the fit. See table 1 for numerical details of the deconvoluted peaks. (* Winfit 1.2- S. Krumm, Institüt für Geologie-Erlangen)

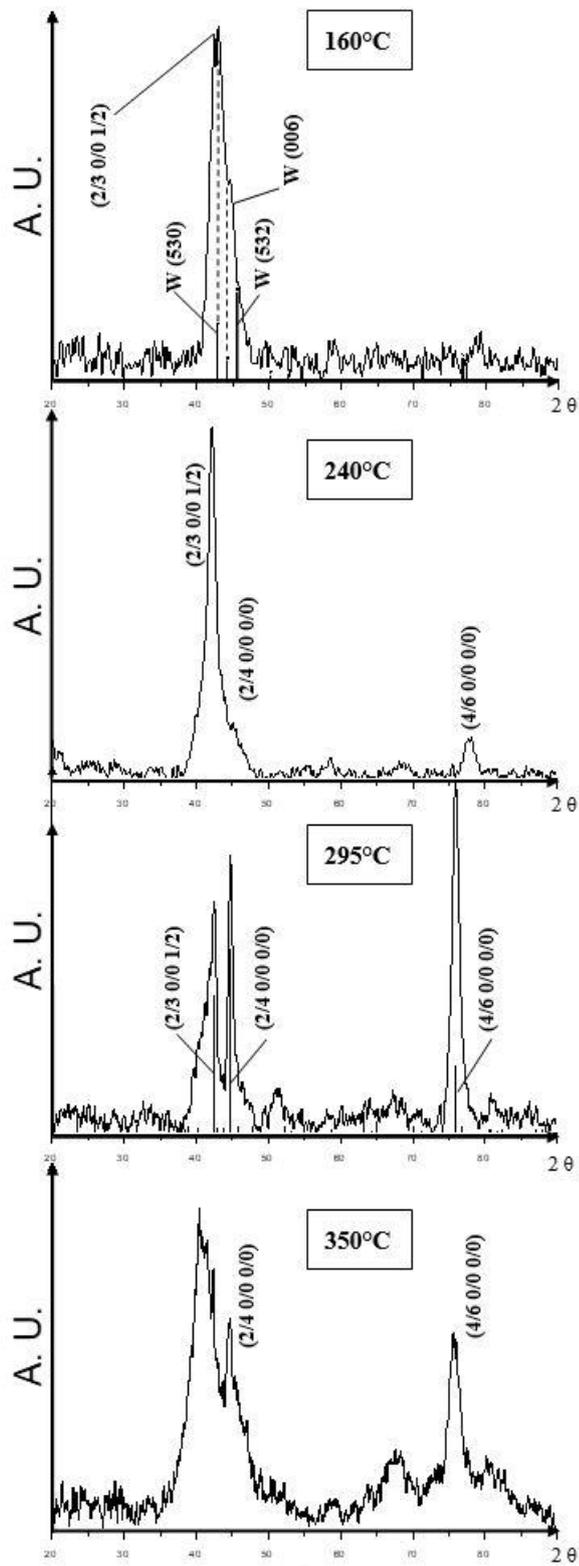

Figure 6 : TEM diffraction pattern of films deposited at (a) 240°C and (b) 295°C.

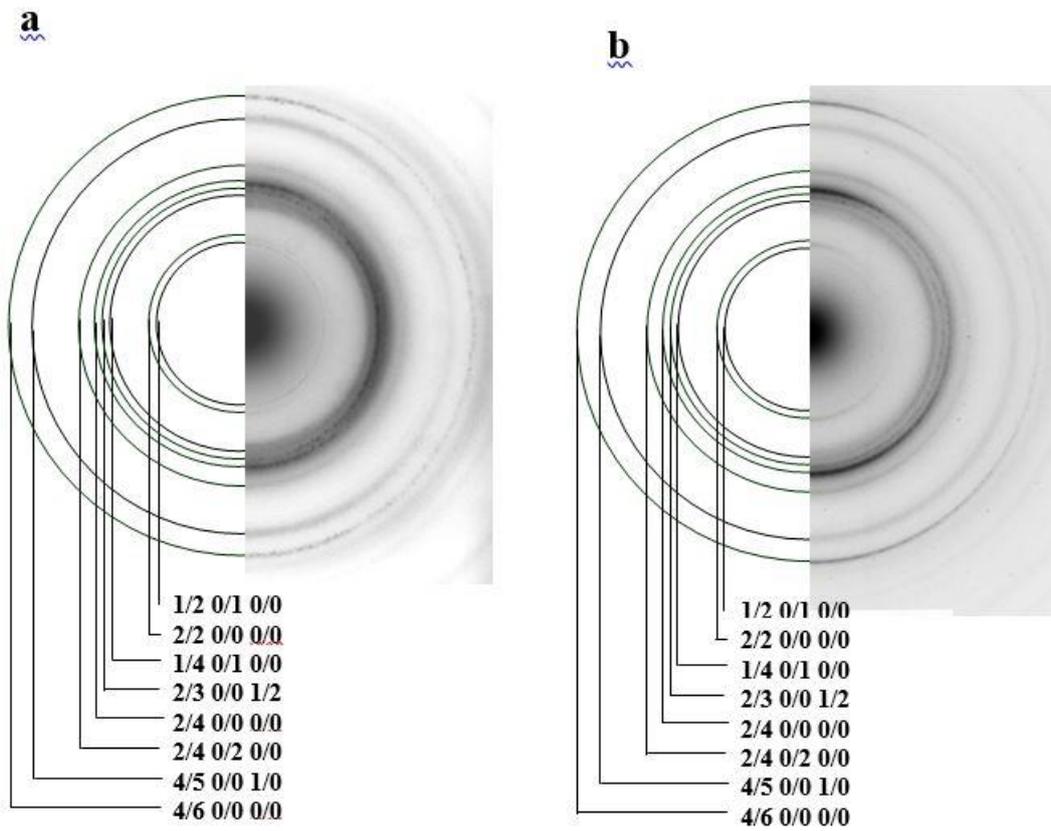

Figure 7 : Electron difraction pattern (a) confirming the presence of a texture in the 295°C film and dark field image (b) made for the measure of the average size of grains of the same film.

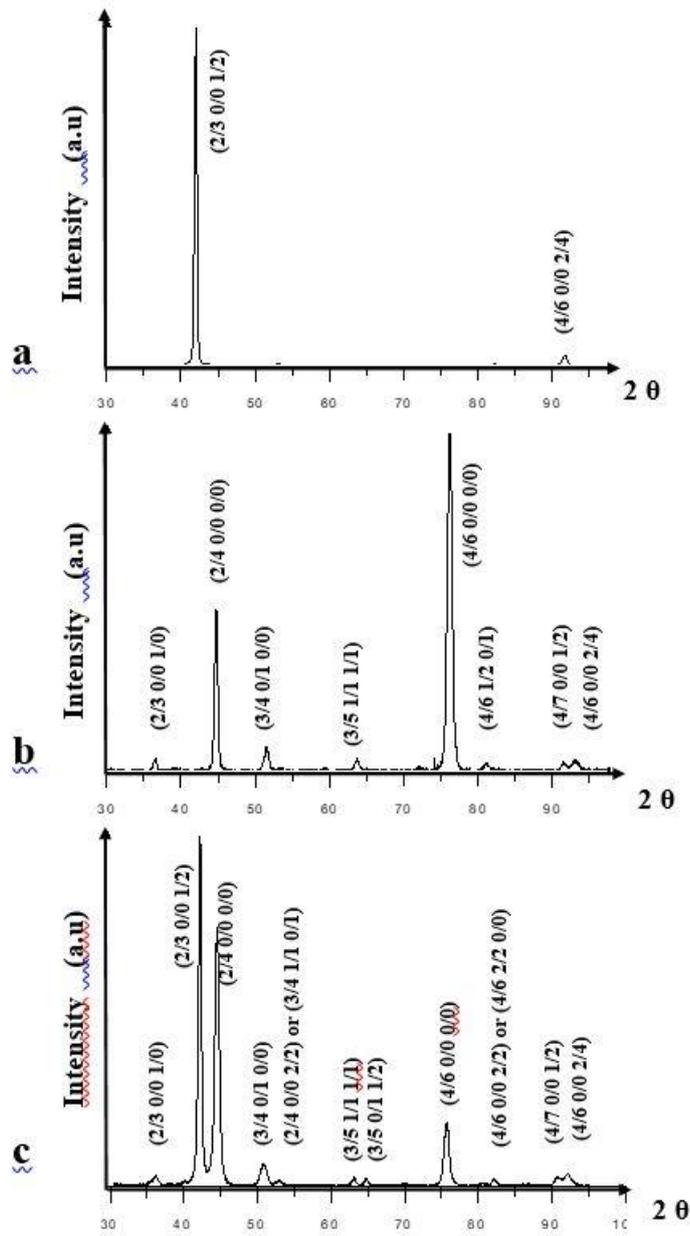

Figure 8 : X ray diffraction patterns performed on a Ti-Ni-Zr film deposited by PLD at 275°C, (a) Bragg-Brentano θ-2θ diffraction pattern, (b) Grazing incidence (0.5°) diffraction pattern, (c) Addition of the 22 scans recorded at different χ (χ =0 to 55°) performed on a four circles diffractometer. (λ = 0.178897 nm).

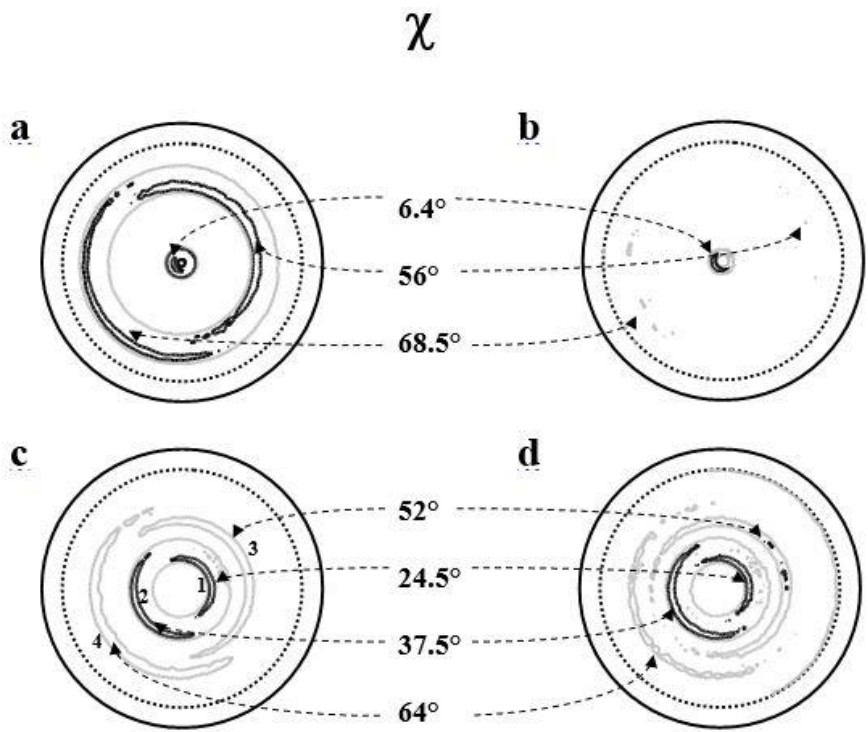

Figure 9 : Pole figures recorded on a Ti-Ni-Zr film deposited by PLD at 275°C on a four circles diffractometer of the
(a) (2/3 0/0 1/2) peak at 2θ = 42°.
(b) (4/6 0/0 2/4) peak at 2θ = 92.3°.
(c) (2/4 0/0 0/0) peak at 2θ = 44°.
(d) (4/6 0/0 0/0) peak at 2θ = 76°.
The dotted circle is the limit of measurements.

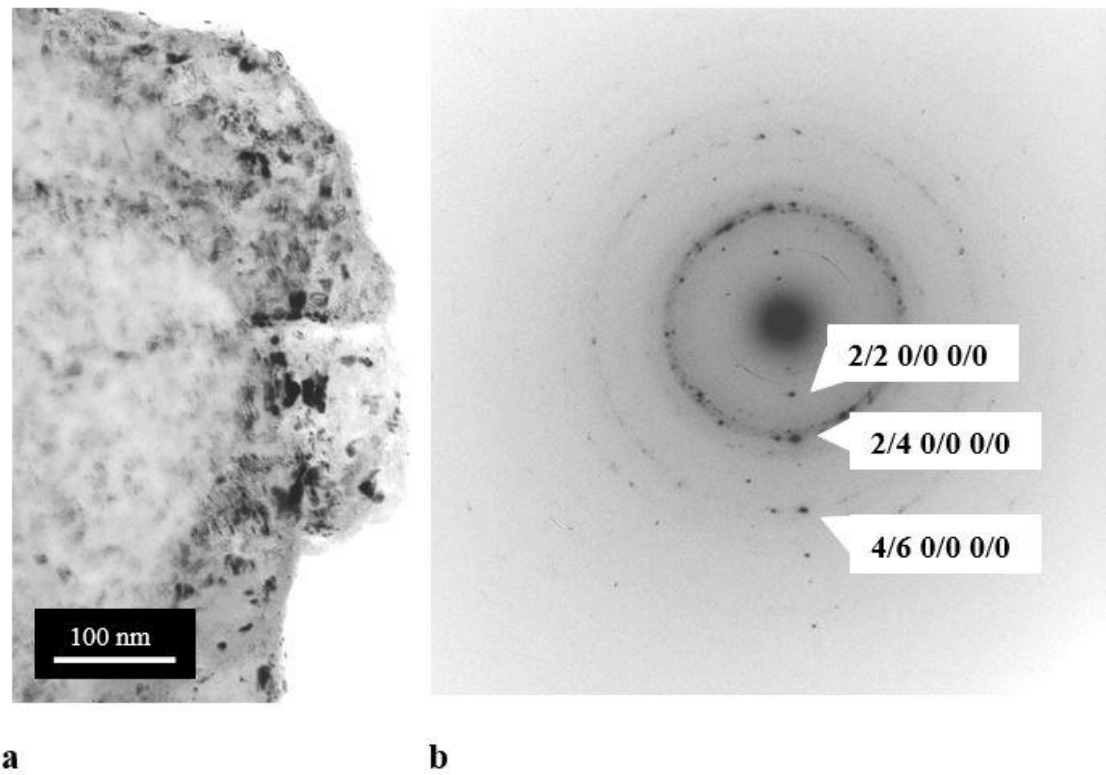

Figure 10 : Film prepared at 275°C : (a) Bright field image showing the typical size of the grains (b) Selected area diffraction pattern of a bigger grain than the average ones observed in the 1 μm film prepared at 275°C.

**Tables**

| $Q_{//}$ nm$^{-1}$ from GXRD (295°C) | $a_{6D}$ nm from GXRD (295°C) ±0.002 nm | $Q_{//}$ nm$^{-1}$ from TEMDP (240°C and 295°C) | $a_{6D}$ nm from TEMDP (240°C and 295°C) ±0.04 nm | $Q_{//}$ nm$^{-1}$ from 4 circles cumulated pattern (275°C) | $a_{6D}$ nm from 4 circles cumulated pattern (275°C) ±0.001 nm | $N/M$ Cahn et al. [18] | Indices |
|---|---|---|---|---|---|---|---|
| - | - | 15.0 | 0.71 | - | - | 6 / 9 | 1/2 0/1 0/0 |
| - | - | 16.7 | 0.73 | - | - | 8 / 12 | 2/2 0/0 0/0 |
| - | - | 24.0 | 0.74 | - | - | 18 / 25 | 1/4 0/1 0/0 |
| 25.45 | 0.741 | 25.7 | 0.73 | 25.24 | 0.746 | 18 / 29 | 2/3 0/0 1/2 |
| 26.74 | 0.740 | 27.2 | 0.73 | 26.57 | 0.745 | 20 / 32 | 2/4 0/0 0/0 |
| - | - | 30.8 | 0.73 | 30.09 | 0.746 | 26 / 41 | 3/4 0/1 0/0 |
| - | - | 39.2 | 0.72 | - | - | 42 / 65 | 4/5 0/0 1/0 |
| 43.24 | 0.741 | 43.9 | 0.73 | 43.09 | 0.743 | 52 / 84 | 4/6 0/0 0/0 |
| - | - | - | - | 49.96 | 0.743 | 70/ 113 | 4/7 0/0 1/2 |
| - | - | - | - | 50.60 | 0.744 | 72 / 116 | 4/6 0/0 2/4 |

Table 1: Lists of the $Q_{//}$ values obtained from the experimental measurements on X-ray or transmission electron diffraction patterns, with their indexation and allowing the calculation of the six dimension lattice parameter $a_{6D}$ of the icosahedral phase (equation 1, see the text).

| $2\theta$ (°) | 42.010 | 42.168 | 45.751 | 77.890 |
|---|---|---|---|---|
| FWHM (°) | 3.493 | 0.693 | 2.464 | 2.8 |
| Indexation | 2/3 0/0 1/2 | 2/3 0/0 1/2 | 2/4 0/0 0/0 | 4/6 0/0 0/0 |
| $a_{6D}$ (nm) | 0.748 | 0.745 | 0.725 | 0.725 |
| Estimation of the size of particles (nm) | ≈ 3 | ≈ 16 | ≈ 4 | ≈ 5 |
| Domain | 2 | 1 | 2 | 2 |

Table 2: Numerical data of peaks of the GXRD pattern of the film prepared at 240°C. The data of the 77.89° peak are directly measured on pattern, the other information was obtained by deconvolution of the 42° peak. See figures 4 and 5 for the pattern and the deconvolution curves.

| $T_s$ (°C) | 160 | 240 | 295 | 350 |
|---|---|---|---|---|
| Size (nm) (miminum-maximum) | 4-8 | 2-10 | 2-19 | 2-10 |

Table 3: Average size of the grains of Ti-Ni-Zr 50 nm thick films versus the substrate temperature $T_s$

**Appendix:**

| Cahn et al. [18] Indexation system | Cahn et al. [18] N/M | Bancel et al. [19] Indexation system | Order of symmetry of the axis of the same indices |
|---|---|---|---|
| 2/3 0/0 1/2 | 18/29 | 1/0 0/0 0/0 | 5 |
| 2/4 0/0 0/0 | 20/32 | 1/1 0/0 0/0 | 2 |
| 3/4 0/1 0/0 | 26/41 | 1/1 1/1 0/1 | 3 |
| 4/6 0/0 0/0 | 52/84 | 1/0 1/0 0/0 | 2 |
| 4/6 0/0 2/4 | 72/116 | 2/0 0/0 0/0 | 5 |

Correspondence between two systems of indexation of the icosahedral structure for the most intense reflections. The symmetry of the axes perpendicular to the indexed planes is given in the right hand side column.